\documentclass[a4paper,11pt]{article}
\pdfoutput=1 

\usepackage{jcappub} 

\usepackage[T1]{fontenc} 

\usepackage{verbatim}
\usepackage[T1]{fontenc}
\usepackage[utf8]{inputenc}
\usepackage[american]{babel}
\usepackage{epsfig}
\usepackage{graphicx}
\usepackage{booktabs}
\usepackage{multirow}
\usepackage{dcolumn}
\usepackage{amsmath}
\usepackage{mathtools}
\usepackage{amsfonts}
\usepackage{amssymb}
\usepackage{epstopdf}
\usepackage{bm}
\usepackage{siunitx}
\usepackage{braket}
\usepackage{enumitem}
\usepackage{soul}
\usepackage[table]{xcolor}
\usepackage{color}
\usepackage{transparent}
\usepackage{pifont}

\definecolor{navyblue}{rgb}{0.0, 0.0, 0.5}
\definecolor{royalblue}{rgb}{0.25, 0.41, 0.88}
\definecolor{cadmiumgreen}{rgb}{0.0, 0.42, 0.24}
\definecolor{blue-violet}{rgb}{0.54, 0.17, 0.89}
\definecolor{darkviolet}{rgb}{0.58, 0.0, 0.83}
\definecolor{orange(colorwheel)}{rgb}{1.0, 0.5, 0.0}

\usepackage{hyperref}
\hypersetup{
    colorlinks=true, 
    linkcolor=royalblue, 
    citecolor=magenta}

\newcommand\ee{\end{equation}}
\newcommand\be{\begin{equation}}
\newcommand\eea{\end{eqnarray}}
\newcommand\bea{\begin{eqnarray}}

\usepackage{booktabs}
\usepackage{multirow}
\usepackage{dcolumn}
\usepackage{colortbl}

\definecolor{magenta(process)}{rgb}{1.0, 0.0, 0.56}

\definecolor{darkspringgreen}{rgb}{0.09, 0.45, 0.27}

\definecolor{royalblue(web)}{rgb}{0.25, 0.41, 0.88}

\title{\boldmath Cosmological constraints in extended parameter space from the Planck 2018 Legacy release}

\author[a]{Eleonora Di Valentino,}
\author[b,1]{Alessandro Melchiorri,\note{Corresponding author.}}
\author[c,d,e]{Joseph Silk}

\affiliation[a]{Jodrell Bank Center for Astrophysics, School of Physics and Astronomy, University of Manchester, Oxford Road, Manchester, M13 9PL, UK}
\affiliation[b]{Physics Department and INFN, Universit\`a di Roma ``La Sapienza'', Ple Aldo Moro 2, 00185, Rome, Italy} 
\affiliation[c]{Institut d'Astrophysique de Paris (UMR7095: CNRS \& UPMC- Sorbonne Universities), F-75014, Paris, France}
\affiliation[d]{Department of Physics and Astronomy, The Johns Hopkins University Homewood Campus, Baltimore, MD 21218, USA}
\affiliation[e]{BIPAC, Department of Physics, University of Oxford, Keble Road, Oxford OX1 3RH, UK}

\emailAdd{eleonora.divalentino@manchester.ac.uk}
\emailAdd{alessandro.melchiorri@roma1.infn.it}
\emailAdd{silk@iap.fr}

\date{\today}

\abstract{We present new constraints on extended cosmological scenarios using the recent data from the Planck 2018 Legacy release. In addition to the $6$ parameters of the standard $\Lambda$CDM model, we also simultaneously  vary  the dark energy equation of state, the neutrino mass, the neutrino effective number, the running of the spectral index and the lensing amplitude $A_L$.
We confirm that a resolution of the Hubble tension is given by a dark energy equation of state with $w<-1$, ruling out quintessence models at high statistical significance. This solution is, however, not supported by BAO and Pantheon data. We find no evidence for evolving dark energy. The neutrino effective number is always in agreement with the expectations of the standard model based on three active neutrinos. The running of the spectral index also is consistent with zero. Despite the increase in the number of parameters, the $A_L$ lensing anomaly is still present at more than two standard deviations. The $A_L$ anomaly significantly affects the bounds on the neutrino mass  that  can be larger by a factor four with respect to those derived under standard $\Lambda$CDM. While CMB lensing data reduces the evidence for $A_L>1$, the inclusion of BAO and Pantheon increases its statistical significance.}

\begin{document}
\maketitle
\flushbottom

\section{Introduction}

The recent Planck 2018 Legacy data release \cite{planck2018} has provided the most accurate measurements of  Cosmic Microwave Background anisotropies to date. Thanks to these measurements, very stringent constraints on several cosmological parameters have been presented. However, those constraints have been obtained under the assumption of a theoretical model. Obviously, for the reliability of the constraints, it is mandatory that the values of the parameters inferred by Planck must be consistent with those derived by independent and complementary observables. While  good agreement is present between Planck and combined analyses of Baryonic Acoustic Oscillations (BAO, hereafter) (see \cite{planck2018})  significant discordance is present in the value of the Hubble constant measured using luminosity distances of Type Ia supernovae. Indeed, while under the assumptions of $\Lambda$CDM,  the Planck dataset provides the value $H_0=67.27\pm0.60$ km/s/Mpc at $68 \%$ C.L. ($H_0=67.67\pm0.45$ km/s/Mpc at $68 \%$ C.L. from Planck+BAO), the recent Riess et al. 2019 result \cite{riess2019} gives $H_0=74.03\pm1.42$ km/s/Mpc at $68 \%$ C.L., i.e. in discordance at the level of $4.4$ standard deviations.
While combined analyses of BAO, Pantheon data, primordial Big Bang Nucleosynthesis, and a conservative Planck bound on the acoustic scale $\theta_{MC}$ gives $H_0=67.9\pm0.8$ km/s/Mpc at $68 \%$ C.L. (see \cite{planck2018}), in very good consistency with the Planck result, recent determinations of $H_0$ from four multiply-imaged quasar systems through strong gravitational lensing made by the H0liCOW collaboration \cite{Wong:2019kwg} have provided $H_0=73.3^{+1.7}_{-1.8}$ km/s/Mpc at $68 \%$ C.L., in good agreement with the Riess et al., 2019 result.

While undetected experimental systematics can still play a role,  perhaps the most  promising one being that of star formation bias \cite{2015ApJ...802...20R},
the increase during the years of the statistical significance in the Hubble tension suggests a crisis for the $\Lambda$CDM cosmological scenario, hinting at the presence of new physics.
In this respect, several physical solutions to the Hubble tension have been already proposed in the literature  (see e.g. \cite{DiValentino:2017rcr,DiValentino:2017iww,Sola:2017jbl,Kumar:2017dnp,Colgain:2018wgk,Poulin:2018cxd,Yang:2018euj,Yang:2018uae,Nunes:2018xbm,Poulin:2018dzj,Capozziello:2018jya,Khosravi:2017hfi,Agrawal:2019lmo,Adhikari:2019fvb,Gelmini:2019deq,Martinelli:2019dau,Pandey:2019plg,Vattis:2019efj,Yang:2019qza,Yang:2019uzo,Yang:2019nhz,Martinelli:2019krf,bahamas}).

The Hubble tension is not, however, the only relevant anomaly from Planck 2018. Another important tension is present in the Planck dataset itself: the Planck CMB angular spectra indeed show a preference for a larger amplitude of the lensing signal with respect to what is expected in $\Lambda$CDM at more than three standard deviations. Indeed, parametrizing the amplitude of CMB lensing by the effective $A_L$ parameter introduced in \cite{alens}, the Planck team has found $A_L=1.18\pm0.14$ at $95 \%$ C.L. \cite{planck2018}, i.e. at odds of about three standard deviations with the $\Lambda$CDM prediction of $A_L=1$. Also in this case, the discordance is puzzling since the  lensing signal obtained again by Planck but in an independent way through measurements of the angular trispectrum is consistent with $\Lambda$CDM \cite{planck2018}. Again, several theoretical solutions have been proposed. The simplest one is to allow a positive curved universe and indeed the Planck CMB spectra do provide evidence for curvature at more than $99 \%$ C.L. \cite{planck2018}. Curvature, however, places the Planck dataset in strong disagreement with BAO and increases the tension significantly with local measurements, Riess et al. 2019 included \cite{diva}. Other possibilities include modified gravity \cite{DiValentino:2015bja}, compensated primordial isocurvature perturbations \cite{Valiviita:2017fbx,Grin:2013uya} and oscillations in the primordial power spectrum \cite{Domenech:2019cyh}. All these modifications, however, are in disagreement with the Planck CMB lensing trispectrum constraint.

In this paper, we follow the method already adopted in \cite{DiValentino:2017zyq,DiValentino:2016hlg,DiValentino:2015ola} 
(but see also \cite{Poulin:2018zxs,Choudhury:2018adz}) by considering a global analysis of current cosmological data but in a significantly more extended cosmological scenario than $\Lambda$CDM. In practice we do not try to solve any single tension with a specific theoretical mechanism, but we allow for a significant number of motivated extensions of $\Lambda$CDM, almost doubling the number of parameters, and looking for a possible combination of parameters that could solve or at least ameliorate, the current discordances. This kind of update is undoubtedly important given the recent Planck 2018 Legacy Release and the new Riess et al. 2019 constraint on $H_0$. 

\section{Method}

The $\Lambda$CDM model is based on the assumption of inflation, cold dark matter, and of a cosmological constant. When compared with CMB observations, there are essentially $6$ independent parameters that can be constrained (see e.g. \cite{planck2018}): the baryon and cold dark matter densities $\Omega_b h^2$ and $\Omega_{c}h^2$, the angular size of the sound horizon at decoupling $\theta_{MC}$, the spectral index $n_s$ and the amplitude $A_S$ of the primordial scalar perturbations, and the optical depth at reionization $\tau$.

Following our previous work, we extend the $\Lambda$CDM model by considering the following additional parameters:

\begin{itemize}
    \item The {\bf running of the spectral} index of inflationary perturbations {\bf $\alpha_s = d n_{s} / d ln k$}. 
    Since it is a dynamical process, running is expected in any inflationary model. For slow-roll inflation generally running
    is predicted at the level of $\sim (1-n_S)^2 \sim 10^{-3}$ (see e.g. \cite{Garcia-Bellido:2014gna}), but it can be larger for several inflationary scenarios (see e.g. \cite{Easther:2006tv,Kohri:2014jma,chung}).
    
    \item The {\bf dark energy equation of state $w$}, assumed either as a constant with redshift $w$, or by introducing a 
    redshift dependence following the CPL form \cite{chevallier,linder}, $w(z)=w+(1-a)w_1$, where $a$ is the adimensional cosmological scale factor normalized to unity today.
    
    \item The {\bf effective number of relativistic particles} at recombination {\bf $N_{eff}$}. This parameter, in the case of three neutrinos species relativistic at recombination, is given by $N_{eff}=3.046$ (see e.g. \cite{Mangano:2005cc,Archidiacono:2011gq,deSalas:2016ztq}) but it can be larger if additional relativistic degrees of freedom (see e.g. \cite{DiValentino:2011sv,DiValentino:2013qma,DiValentino:2015wba,Baumann:2016wac,Gariazzo:2015rra,Smith:2006nka,Dimastrogiovanni:2017tvd, DiValentino:2017oaw}) are  present at that epoch.
    
    \item The sum of neutrino masses $\Sigma m_{\nu}$. Current neutrino oscillation experiments have provided conclusive evidence that neutrinos are massive by measuring the mass differences between neutrino flavours. The total mass is however still unknown. Current laboratory experiments place a lower limit of $\Sigma m_{nu} >0.05$ eV (see e.g. \cite{Capozzi:2017ipn}).
    
    \item The amplitude of the dark matter lensing contribution to the CMB angular power spectra $A_{L}$ \cite{alens}. The recent Planck Legacy data release shows evidence for $A_L>1$ at about three standard deviation. It is therefore important to also consider this parameter. We remind the reader, however, that it is only an {\it effective} and, ultimately, unphysical parameter.
\end{itemize}

We consider the following cases of increasing numbers of parameter: $\Lambda$CDM+$w$+$\alpha_S$+$N_{eff}$+$\Sigma m_{\nu}$ ($10$ Parameters, $\Lambda$CDM+$w$+$\alpha_S$+$N_{eff}$+$\Sigma m_{\nu}$+$A_L$ ($11$ parameters), $\Lambda$CDM+$w$+$w_a$+$\alpha_S$+$N_{eff}$+$\Sigma m_{\nu}$+$A_L$ ($12$ parameters).

These models are compared with the following datasets:

\begin{itemize}
    \item The Planck $2018$ temperature and polarization CMB angular power spectra. This corresponds to the the Planck TT,TE,EE+low E dataset used in \cite{planck2018}. We refer to this dataset simply as {\bf Planck}.
    \item The Planck constraints on the CMB lensing potential obtained from a trispectrum analysis of temperature and polarization CMB maps \cite{Aghanim:2018oex}. We refer to this dataset as {\bf Lensing}.
    \item The Baryon Acoustic Oscillation data from the compilation used in \cite{planck2018}. This consist of data from the
     6dFGS~\cite{6dFGS}, SDSS MGS~\cite{mgs}, and BOSS DR12~\cite{bossdr12} surveys. We refer to this dataset as {\bf BAO}.
    \item The luminosity distance data of type Ia supernovae from the PANTHEON catalog \cite{pantheon}. We refer to this dataset as {\bf Pantheon}.
    \item The most recent determination of the Hubble constant from Riess et al. 2019. This is assumed as a gaussian prior on the Hubble constant of $H_0=74.03\pm1.42$ km/s/Mpc. We refer to this prior as {\bf R19} \cite{riess2019}.
    \item The cosmic shear data from the first release of the Dark Energy Survey {\bf DES} \cite{DES}.
\end{itemize}

The comparison between theory and data is made using the recently released Planck 2019 likelihood \cite{plancklike} adopting 
the public available CosmoMC code based \cite{Lewis&Bridle} on a Monte Carlo Markov chain algorithm. The theoretical predictions are made using the CAMB Boltzmann integrator \cite{Lewis:1999bs}.

\section{Results}

\subsection{$10$ parameters model: $\Lambda$CDM+$w$+$\alpha_S$+$N_{\rm eff}$+$\Sigma m_{\nu}$}

\begin{center}      
\begin{table*}      
\scalebox{0.8}{
\begin{tabular}{cccccccccccccccc}       
\hline\hline                                                                                                                    
Parameters & Planck   & Planck & Planck& Planck & Planck & \\ 
 &  &+R19 & +lensing  & +BAO & + Pantheon \\ \hline
 
 $\Omega_b h^2$ & $    0.02217 \pm 0.00024$ &  $    0.02216\pm 0.00024$ & $    0.02218\pm 0.00024$ & $    0.02224\pm 0.00022$ & $    0.02212\pm 0.00023$ \\
 
$\Omega_c h^2$ & $    0.1163\pm 0.0033$  & $    0.1165\pm 0.0034$ & $    0.1159\pm 0.0032$ & $    0.1165\pm 0.0033 $& $    0.1162\pm 0.0033 $\\

$100\theta_{\rm MC}$ & $    1.04135\pm 0.00048$ &  $    1.04132\pm 0.00050$ &  $    1.04137\pm 0.00049$ & $    1.04136\pm 0.00049$ & $    1.04138\pm 0.00051$\\

$\tau$ & $    0.0546\pm 0.0080$ &  $    0.0556^{+0.0074}_{-0.0093}$ &  $    0.0531\pm 0.0078$ & $    0.0558\pm 0.0082$ & $    0.0548\pm 0.0081$\\

$\Sigma m_{\nu}$ [eV] & $    <0.328$ &  $    <0.320$ & $    <0.368$ & $    <0.167$ & $    <0.298$\\

$w$ & $    -1.64^{+0.28}_{-0.40}$ &  $    -1.34^{+0.12}_{-0.09}$ & $    -1.65^{+0.30}_{-0.39}$ & $    -1.072^{+0.079}_{- 0.061}$ & $    -1.064^{+0.048}_{-0.040}$\\

$N_{\rm eff}$ & $    2.76\pm 0.22$ &  $    2.76^{+0.22}_{-0.25}$ & $    2.76^{+0.21}_{-0.23}$ & $    2.81\pm 0.22$ & $    2.73\pm 0.22$\\

${\rm{ln}}(10^{10} A_s)$ & $    3.036\pm 0.019$ &  $    3.039^{+0.017}_{-0.023}$ & $    3.032\pm 0.018$ & $    3.040\pm 0.019$ & $    3.037\pm 0.019$\\

$n_s$ & $    0.951\pm 0.011$ & $    0.951\pm 0.011$ &   $    0.952\pm 0.011$ & $    0.954\pm 0.010$ &  $    0.949\pm 0.011$\\

$\alpha_S$ & $    -0.0098\pm 0.0079$ &  $    -0.0104\pm 0.0078$ & $    -0.0087\pm 0.0081$ & $    -0.0100\pm 0.0077$ & $    -0.0117\pm 0.0081$\\

$H_0 $[km/s/Mpc] & $   >65.2$&  $   74.2\pm 1.4$ & $   >65.7$ & $   67.9\pm 1.7$ & $   66.3\pm 1.7$\\

$\sigma_8$ & $    0.95^{+0.11}_{-0.06}$ &  $    0.874^{+0.024}_{-0.020}$ & $    0.94^{+0.10}_{-0.06}$ & $    0.821\pm 0.020$ & $    0.806^{+0.024}_{-0.016}$\\

$S_8$ & $    0.774^{+0.031}_{-0.042}$ &  $    0.805\pm 0.020$ & $  0.765^{+0.028}_{-0.042}  $ & $    0.824^{+0.015}_{-0.014}$ & $    0.829\pm 0.017$\\

\hline\hline                                                  
\end{tabular}  
}
\caption{Constraints at 68$\%$~CL errors on the cosmological parameters in case of the $10$ parameters model, $\Lambda$CDM+$w$+$\alpha_S$+$N_{eff}$+$\Sigma m_{\nu}$, using different combinations of datasets. The quoted upper/lower limits are at 95$\%$~CL.}
\label{10p}                                              
\end{table*}
\end{center}

\begin{figure*}[!hbtp]
\begin{center}
\includegraphics[width=.3\textwidth,keepaspectratio]{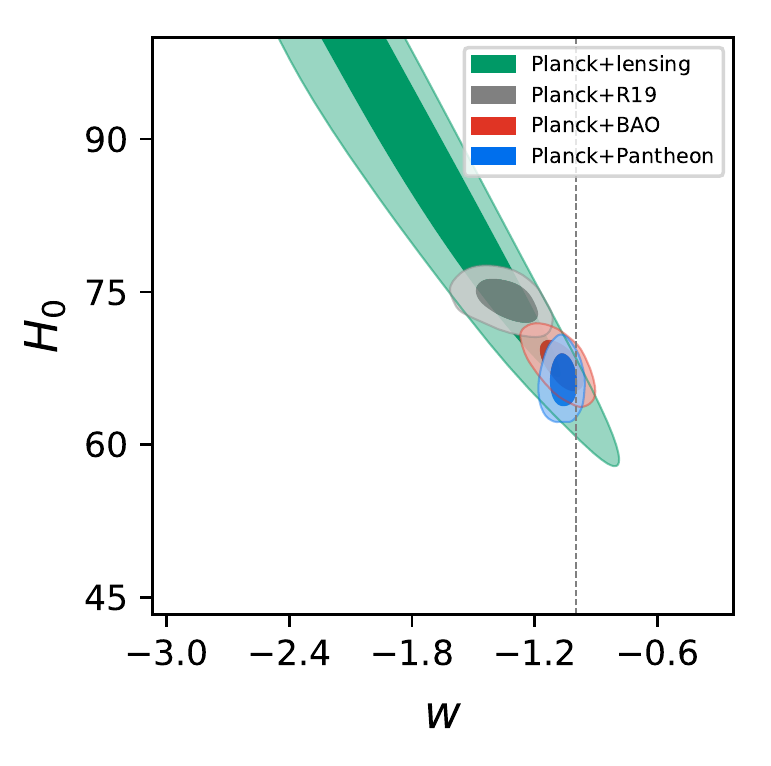}
\includegraphics[width=.3\textwidth,keepaspectratio]{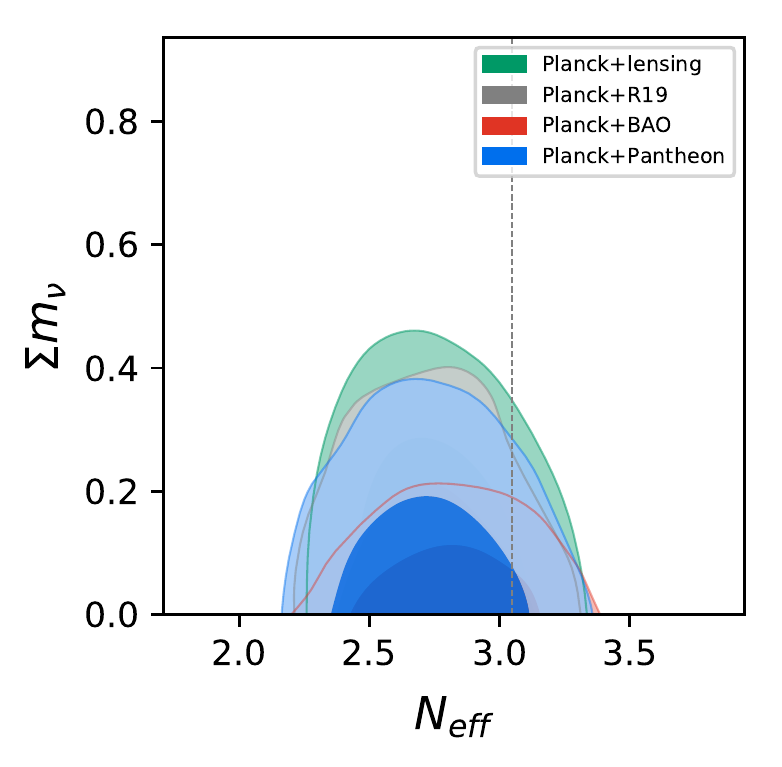}
\includegraphics[width=.31\textwidth,keepaspectratio]{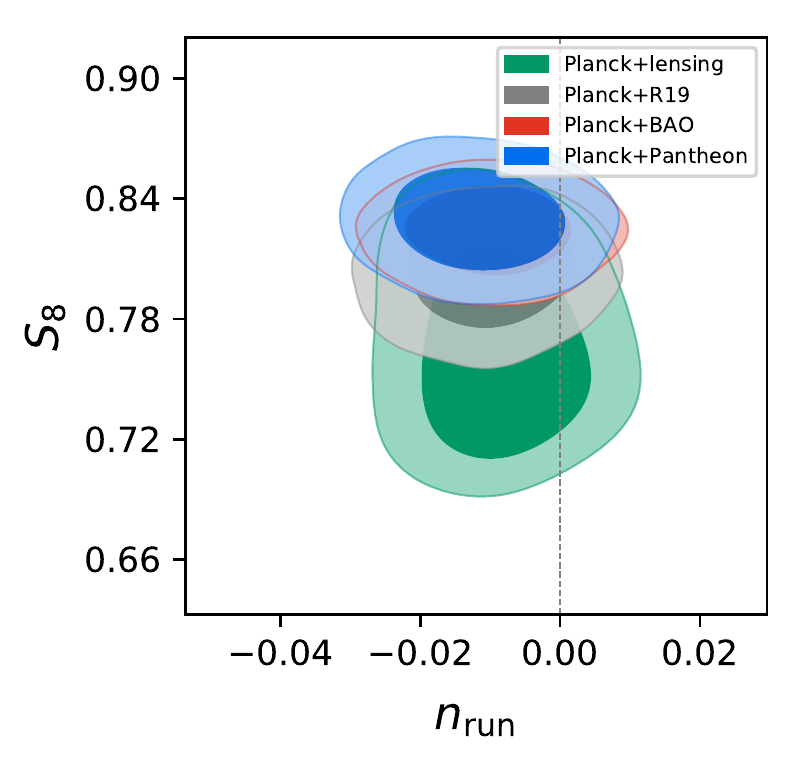}
\includegraphics[width=.31\textwidth,keepaspectratio]{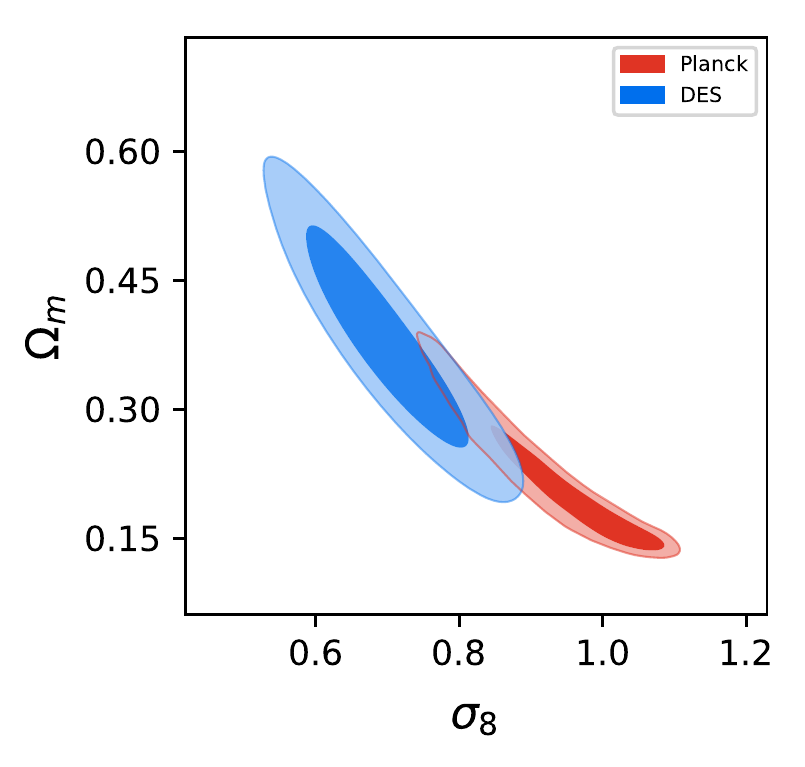}

\end{center}
\caption{Constraints at the $68\%$ and $95\%$ C.L. on the $w$ vs $H_0$, $\Sigma m_{\nu}$ vs $N_{eff}$, and $S_8$ vs $\alpha_s$ planes for the Planck+Lensing, Planck+R19, Planck+BAO, and Planck+Pantheon datasets (Top). The Planck and DES constraints at the $68\%$ and $95\%$ C.L. on the $\sigma_8$ vs $\Omega_m$ plane are also showed (Bottom). A $10$ parameters model, $\Lambda$CDM+$w$+$\alpha_S$+$N_{eff}$+$\Sigma m_{\nu}$, is assumed in the analysis.}
\label{fig10p}
\end{figure*}

The constraints on the $10$ parameters of this extended scenario are reported in Table~\ref{10p}, while on the top of Figure~\ref{fig10p} we show the 2D constraints at the $68\%$ and $95\%$ C.L. on the $w$ vs $H_0$, $\Sigma m_{\nu}$ vs $N_{eff}$, and $S_8$ vs $\alpha_s$ planes for the Planck+Lensing, Planck+R19, Planck+BAO, and Planck+Pantheon datasets, respectively. In the bottom of Figure~\ref{fig10p} we show the Planck and DES constraints at the $68\%$ and $95\%$ C.L. on the $\sigma_8$ vs $\Omega_m$ plane.
We can derive the following conclusions from this analysis:

\begin{itemize}
    \item Both the Planck and Planck+lensing datasets are unable, in our extended parameter framework, to place constraints on the Hubble constant $H_0$. As we can also see from the top left panel of Figure~\ref{fig10p}, a degeneracy is present with the dark energy equation of state that does not allow any strong constraint on $H_0$ and $w$. This clearly reinforces the statement that the current tension between the Planck and R19 values of $H_0$ is based on the assumptions of the $\Lambda$CDM model. Once this model is extended, the two datasets can be easily put into agreement.
    \item For the Planck+$R19$ dataset,  all the parameters are consistent with the expectations of the standard $\Lambda$CDM model with the most notable exception of $w$, found in this case to be $w=-1.34^{+0.12}_{-0.09}$ at $68 \%$ C.L. ($w=-1.34^{+0.20}_{-0.22}$ at $95 \%$ C.L.), i.e. less than $-1$ at about three standard deviations. A similar conclusion is reached for the Planck+lensing+R19 dataset where we obtain $w=-1.35^{+0.22}_{-0.24}$ at $95 \%$ C.L..
    \item Both Planck+BAO and Planck+Pantheon datasets are in perfect agreement with the main expectations of standard $\Lambda$CDM of $w=-1$. Consequently, the bounds on $H_0$ and $w$ derived using the Planck+$R19$ dataset are in significant tension with the corresponding constraints obtained from Planck+BAO or Planck+Pantheon. The value of the Hubble constant from Planck+$R19$ is indeed discordant at $2.9$ standard deviations from the one derived by Planck+BAO and at $3.6$ standard deviations from the value obtained by Planck+Pantheon. At the same time, the equation of state $w$ from Planck+$R19$ is in tension at the level of $2$ standard deviations with the Planck+BAO constraint and at $2.2$ standard deviation with Planck+Pantheon.
    \item The value of the relativistic number of effective neutrinos $N_{eff}$ is in good agreement with the expectations of the standard scenario, even if there is a hint for $N_{eff}<3.04$ at about $1.5$ standard deviations. 
    \item There is an indication slightly above one standard deviation for a negative running of the spectral index in all   of the datasets considered. This, together with the one sigma indication for $N_{eff}<3.04$, brings the value of the spectral index to $n_s\sim0.95$, i.e. about one sigma lower than the $\Lambda$CDM result of $n_s=0.966\pm0.0042$ at $68 \%$ C.L..
    \item  There is no evidence for a neutrino mass at more than two standard deviations in all the datasets. However, the constraints on the neutrino mass are relaxed in the extended scenario with respect to standard $\Lambda$CDM. For example, the conservative Planck+lensing dataset  provides an upper limit of $\Sigma m_{\nu}<0.368$ eV at $95\%$ c.l. to be compared with the upper limit of $\Sigma m_{\nu}<0.24$ eV at $95\%$ c.l. obtained assuming standard $\Lambda$CDM. In the case of Planck+BAO we find  $\Sigma m_{\nu}<0.167$ eV at $95\%$ c.l. to be compared with $\Sigma m_{\nu}<0.126$ eV at $95\%$ c.l. under $\Lambda$CDM.
    \item Values of the $S_8=\sigma_8(\Omega_m/0.3)^{0.5}$ parameter around $\sim 0.78$, as suggested by cosmic shear surveys such as DES \cite{DES} are compatible with the Planck+lensing and Planck+R19 datasets. However both Planck+BAO and Planck+Pantheon prefer higher values of $S_8$ (see Right Panel of Figure~\ref{fig10p}). 
    \item The Planck constraint in the $\sigma_8$ vs $\Omega_m$ plane is in better consistency with the DES constraint than in the $\Lambda$CDM (see Bottom Panel of Figure~\ref{fig10p}).
\end{itemize}

\subsection{$11$ parameters model: $\Lambda$CDM+$w$+$\alpha_S$+$N_{eff}$+$\Sigma m_{\nu}$+$A_L$}

\begin{center}                              
\begin{table*}   
\scalebox{0.8}{
\begin{tabular}{cccccccccccccccc}       
\hline\hline                                                                                                                    
Parameters & Planck   & Planck & Planck& Planck & Planck & \\ 
 &  &+R19 & +lensing  & +BAO & + Pantheon \\ \hline
 
 $\Omega_b h^2$ & $    0.02246 \pm 0.00028$ &  $    0.02248^{+0.00028}_{-0.00032}$ & $    0.02228\pm 0.00026$ & $    0.02264\pm 0.00026$ & $    0.02250\pm 0.00028$ \\
 
$\Omega_c h^2$ & $    0.1172\pm 0.0033$  & $    0.1174\pm 0.0035$ & $    0.1164\pm 0.0033$ & $    0.1175\pm 0.0033 $& $    0.1174^{+0.0031}_{-0.0035} $\\

$100\theta_{\rm MC}$ & $    1.04112\pm 0.00051$ &  $    1.04111\pm 0.00052$ &  $    1.04119\pm 0.00050$ & $    1.04120\pm 0.00049$ & $    1.04111\pm 0.00050$\\

$\tau$ & $    0.0496\pm 0.0086$ &  $    0.0508\pm 0.0091$ &  $    0.0494^{+0.0086}_{-0.0076}$ & $    0.0502\pm 0.0087$ & $    0.0499^{+0.0086}_{-0.0078}$\\

$\Sigma m_{\nu}$ [eV] & $    <0.863$ &  $    <0.821$ & $    <0.714$ & $    <0.352$ & $    <0.822$\\

$w$ & $    -1.27\pm 0.53$ &  $    -1.33^{+0.17}_{-0.11}$ & $    -1.33\pm0.52$ & $    -1.009^{+0.092}_{- 0.070}$ & $    -1.071^{+0.073}_{-0.050}$\\

$N_{\rm eff}$ & $    2.95\pm 0.24$ &  $    2.97\pm 0.26$ & $    2.85\pm 0.23$ & $    3.04\pm 0.23$ & $    2.98^{+0.23}_{-0.25}$\\

$A_L$ & $    1.25^{+0.09}_{-0.14}$ &  $    1.21^{+0.09}_{-0.10}$ & $    1.116^{+0.061}_{-0.096}$ & $    1.213^{+0.076}_{-0.088}$ & $    1.232\pm 0.090$\\

${\rm{ln}}(10^{10} A_s)$ & $    3.027\pm 0.020$ &  $    3.030\pm 0.022$ & $    3.024\pm 0.020$ & $    3.030\pm 0.020$ & $    3.028^{+0.020}_{-0.018}$\\

$n_s$ & $    0.964\pm 0.012$ & $    0.965\pm 0.013$ &   $    0.958\pm 0.012$ & $    0.971\pm 0.012$ &  $    0.965\pm 0.012$\\

$\alpha_S$ & $    -0.0053\pm 0.0085$ &  $    -0.0047\pm 0.0082$ & $    -0.0066\pm 0.0082$ & $    -0.0041\pm 0.0081$ & $    -0.0049\pm 0.0086$\\

$H_0 $[km/s/Mpc] & $   73^{+10}_{-20}$&  $   74.0\pm 1.4$ & $   74^{+10}_{-20}$ & $   67.9\pm 1.7$ & $   66.9\pm 2.0$\\

$\sigma_8$ & $    0.79^{+0.15}_{-0.13}$ &  $    0.811^{+0.051}_{-0.035}$ & $    0.80^{+0.15}_{-0.13}$ & $    0.782\pm 0.025$ & $    0.750^{+0.055}_{-0.034}$\\

$S_8$ & $    0.754^{+0.053}_{-0.041}$ &  $    0.758^{+0.039}_{-0.027}$ & $  0.757^{+0.047}_{-0.038}  $ & $    0.791^{+0.025}_{-0.019}$ & $    0.775^{+0.036}_{-0.026}$\\

\hline\hline                                                  
\end{tabular} 
}
\caption{Constraints at 68$\%$~CL errors on the cosmological parameters in case of the $11$ parameters model using different combinations of the datasets. The quoted upper/lower limits are at 95$\%$~CL.}
\label{tab11p}                                              
\end{table*}                                                    
\end{center} 

\begin{figure*}[!hbtp]
\begin{center}
\includegraphics[width=.31\textwidth,keepaspectratio]{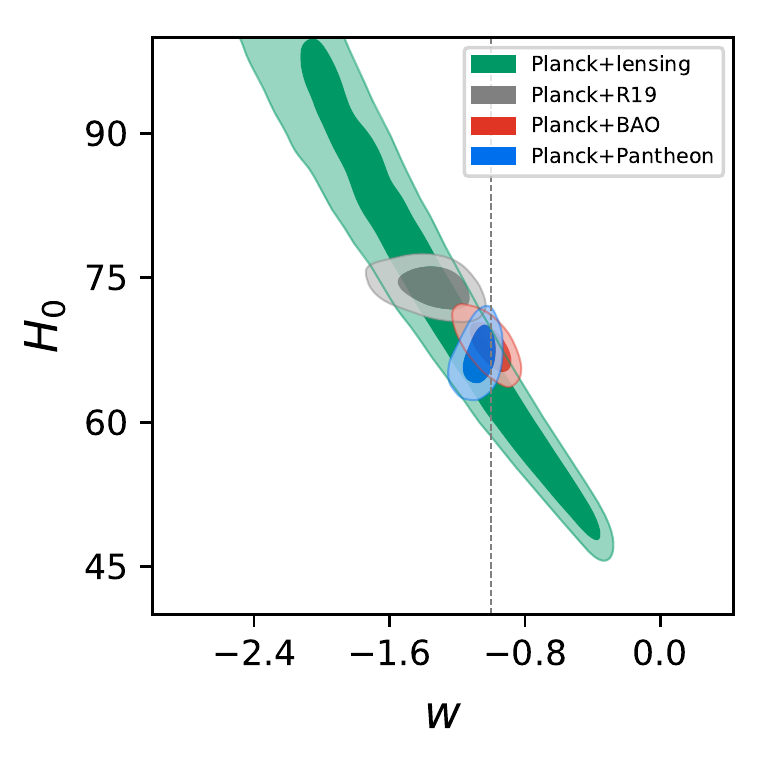}
\includegraphics[width=.31\textwidth,keepaspectratio]{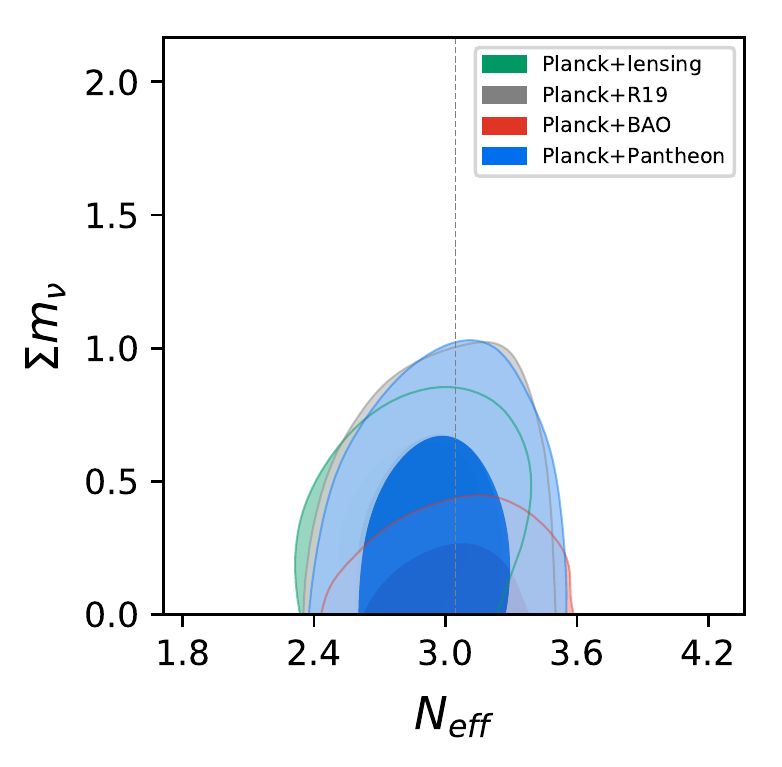}
\includegraphics[width=.31\textwidth,keepaspectratio]{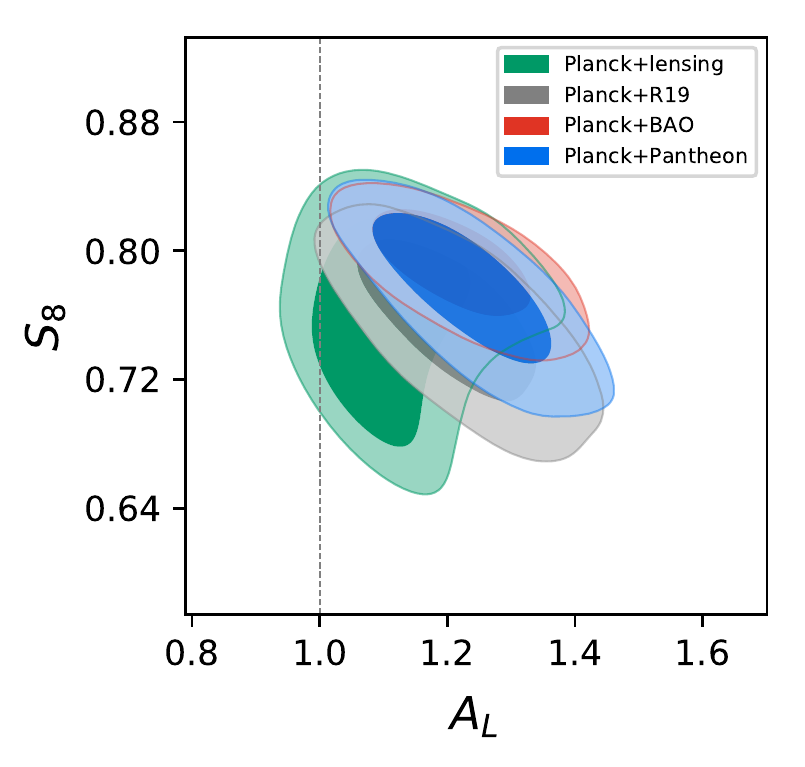}
\includegraphics[width=.31\textwidth,keepaspectratio]{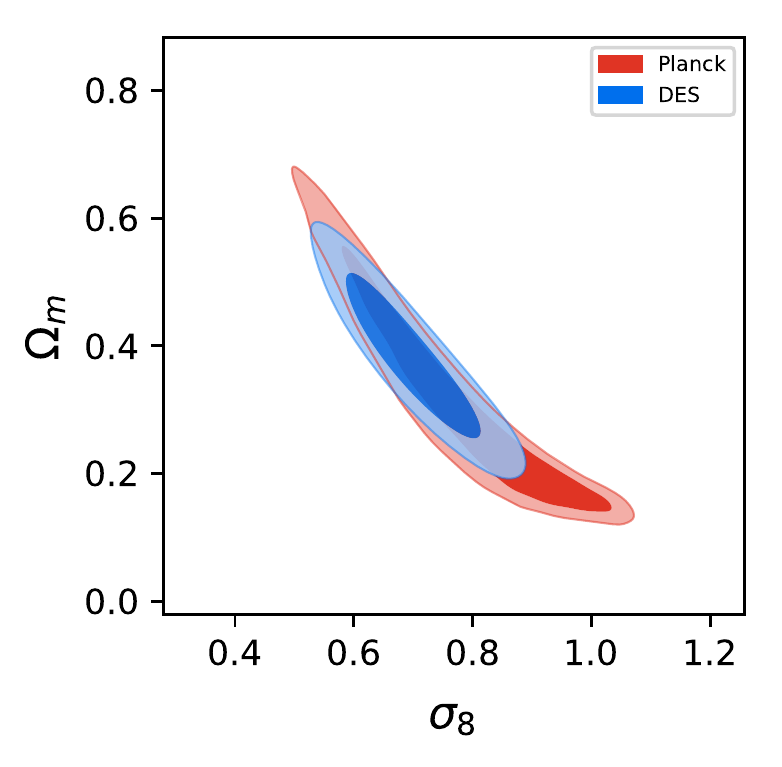}
\end{center}
\caption{Top:constraints at the $68\%$ and $95\%$ C.L. on the $w$ vs $H_0$, $\Sigma m_{\nu}$ vs $N_{eff}$, and $S_8$ vs $A_L$ planes for the Planck+Lensing, Planck+R19, Planck+BAO, and Planck+Pantheon datasets. Bottom: constraints at the $68\%$ and $95\%$ C.L. on the $\sigma_8$ vs $\Omega_m$ plane from Planck and DES, separately.A $11$ parameters model, $\Lambda$CDM+$w$+$\alpha_S$+$N_{eff}$+$\Sigma m_{\nu}$+$A_L$, is assumed in the analysis.}
\label{fig11p}
\end{figure*}

The constraints in the case of $11$ parameters model are reported in Table~\ref{tab11p} for the Planck, Planck+lensing, Planck+R19, Planck+BAO, and Planck+Pantheon datasets. In the Top of Figure~\ref{fig11p}, we plot the 2-D constraints at the $68\%$ and $95\%$ C.L. on the $w$ vs $H_0$, $\Sigma m_{\nu}$ vs $N_{eff}$, and $S_8$ vs $A_L$ planes for the Planck+Lensing, Planck+R19, Planck+BAO, and Planck+Pantheon datasets. In the bottom of Figure~\ref{fig11p} we show the constraints at the $68\%$ and $95\%$ C.L. on the $\sigma_8$ vs $\Omega_m$ plane from Planck and DES, separately.
We can derive the following conclusions:

\begin{itemize}
    \item In all datasets, with the exception of the Planck+lensing dataset, there is a strong indication of an anomalous value of the lensing amplitude with $A_L>1$ at about three standard deviations (we found $A_L=1.25^{+0.45}_{-0.25}$ at $99 \%$ C.L. from Planck alone). The Planck+BAO dataset further increases the indication of the $A_L$ anomaly with $A_L=1.21^{+0.23}_{-0.20}$ at $99 \%$ C.L.. As we can see from the right panel of Figure~\ref{fig11p}, a value of $A_L>1$ shifts all the constraints on the $S_8$ parameter to values more consistent with those recently determined by the KiDS-450 cosmic shear survey~\cite{DiValentino:2018gcu} under $\Lambda$CDM. 
    The Planck+BAO and Planck+Pantheon datasets, for example, are now consistent with the value of $S_8\sim 0.77$, providing $A_L>1$.
    \item As in the $10$ parameters case, the Planck+$R19$ dataset shows evidence for $w<-1$, with $w=-1.21\pm0.18$ at $95 \%$ C.L., i.e. $w<-1$ at about three standard deviations (we found $w=-1.37^{+0.24}_{-0.27}$ at $95 \%$ C.L. for the Planck+lensing+R19 dataset). The Planck+BAO and Planck+Pantheon datasets are still in perfect agreement with the main expectations of standard $\Lambda$CDM of $w=-1$.
    \item The Planck+$R19$ dataset is still in significant tension with the Planck+BAO and the Planck+Pantheon datasets for the values of $H_0$ and $w$. Considering $H_0$, the Planck+$R19$ constraint is in tension at $2.8$ standard deviations with the Planck+BAO bound and at $2.9$ standard deviations with the Planck+Pantheon value. In the case of $w$ the tension with Planck+$R19$ is at $1.7$ standard deviations with Planck+BAO and at $1.5$ standard deviations with Planck+Pantheon. The tension between these datasets is therefore mitigated by the introduction of the $A_L$ parameter with respect to the previous $10$ parameters model.
    \item The value of the relativistic number of effective neutrinos $N_{eff}$ is in perfect agreement with the standard value $N_{eff}=3.046$ and there is no indication of any negative running of the spectral index. The indications of $N_{eff}<3.046$ and for $\alpha_s$, slightly above one standard deviation that we see in the $10$ parameter scenario, simply vanish when the $A_L$ parameter is introduced. At the same time, the constraints on $n_s$ are in complete agreement with those inferred under $\Lambda$CDM.
    \item Unsurprisingly, the constraints on the neutrino mass are significantly relaxed not only with respect to the LCDM scenario but also in comparison to the previous $10$ parameter model. The upper limits on the neutrino mass are about a factor two weaker than those derived in the $10$ parameter case. For example, the Planck+BAO limit is now $\Sigma m_{\nu}<0.352$ eV at $95 \%$ C.L., 
    a factor $\sim 2.1$ weaker than the corresponding bound obtained in the $10$ parameter scenario with the same datasets.
    There is no indication for a neutrino mass at more than two standard deviations in any of the datasets. In any case, since the introduction of $A_L$ affects the constraints on the neutrino mass so strongly, it is of utmost importance to understand the nature of the $A_L$ anomaly before considering current constraints on neutrino mass to be fully reliable.
    \item The constraints in the $S_8$ vs $\Omega_m$ plane from Planck and DES overlap completely suggesting that the tension between these two dataset is fully removed in this $11$ parameter space.
    \end{itemize}

\subsection{$12$ parameters model: $\Lambda$CDM+$w$+$w_a$+$\alpha_S$+$N_{eff}$+$\Sigma m_{\nu}$+$A_L$}
\begin{center}                              
\begin{table*} 
\scalebox{0.8}{
\begin{tabular}{cccccccccccccccc}       
\hline\hline                                                                                                                    
Parameters & Planck   & Planck & Planck& Planck & Planck & \\ 
 &  &+R19 & +lensing  & +BAO & + Pantheon \\ \hline
 
 $\Omega_b h^2$ & $    0.02245 \pm 0.00027$ &  $    0.02246\pm 0.00027$ & $    0.02228\pm 0.00026$ & $    0.02255\pm 0.00026$ & $    0.02245\pm 0.00027$ \\
 
$\Omega_c h^2$ & $    0.1173\pm 0.0035$  & $    0.1173\pm 0.0033$ & $    0.1163\pm 0.0033$ & $    0.1172\pm 0.0035 $& $    0.1174\pm0.0034 $\\

$100\theta_{\rm MC}$ & $    1.04113\pm 0.00052$ &  $    1.04113\pm 0.00050$ &  $    1.04123\pm 0.00052$ & $    1.04118\pm 0.00051$ & $    1.04107\pm 0.00052$\\

$\tau$ & $    0.0500\pm 0.0087$ &  $    0.0516^{+0.0083}_{-0.0097}$ &  $    0.0498\pm 0.0085$ & $    0.0504\pm 0.0085$ & $    0.0497^{+0.0088}_{-0.0077}$\\

$\Sigma m_{\nu}$ [eV] & $    <0.906$ &  $    <0.857$ & $    0.35^{+0.16}_{-0.26}$ & $    <0.515$ & $    <0.928$\\

$w$ & $    -1.02^{+0.67}_{-0.96}$ &  $    -1.22\pm 0.33$ & $    -1.03^{+0.62}_{-0.26}$ & $    -0.62^{+0.30}_{-0.26}$ & $    -1.02\pm 0.16$\\

$w_a$ & $    unconstrained$ &  $    unconstrained$ & $    unconstrained$ & $    -1.29\pm 0.87$ & $    -0.39^{+0.98}_{-0.70}$\\

$N_{\rm eff}$ & $    2.95\pm 0.24$ &  $    2.95\pm 0.24$ & $    2.85\pm 0.23$ & $    2.99\pm 0.24$ & $    2.96\pm 0.24$\\

$A_L$ & $    1.23^{+0.09}_{-0.12}$ &  $    1.21^{+0.09}_{-0.10}$ & $    1.106^{+0.059}_{-0.089}$ & $    1.203^{+0.080}_{-0.089}$ & $    1.236^{+0.087}_{-0.098}$\\

${\rm{ln}}(10^{10} A_s)$ & $    3.028\pm 0.020$ &  $    3.031^{+0.019}_{-0.022}$ & $    3.024\pm 0.020$ & $    3.029\pm 0.019$ & $    3.027\pm 0.020$\\

$n_s$ & $    0.964\pm 0.012$ & $    0.964\pm 0.012$ &   $    0.958\pm 0.012$ & $    0.967\pm 0.012$ &  $    0.964\pm 0.012$\\

$\alpha_S$ & $    -0.0053\pm 0.0087$ &  $    -0.0053\pm 0.0086$ & $    -0.0067\pm 0.0084$ & $    -0.0054\pm 0.0085$ & $    -0.0052\pm 0.0085$\\

$H_0 $[km/s/Mpc] & $   72\pm20$&  $   74.0\pm 1.4$ & $   72\pm20$ & $   64.8^{+2.5}_{-2.9}$ & $   66.8\pm 2.1$\\

$\sigma_8$ & $    0.78^{+0.15}_{-0.14}$ &  $    0.811^{+0.053}_{-0.035}$ & $    0.79^{+0.17}_{-0.12}$ & $    0.751\pm 0.033$ & $    0.745^{+0.056}_{-0.043}$\\

$S_8$ & $    0.760\pm 0.054$ &  $    0.758^{+0.040}_{-0.029}$ & $  0.766\pm 0.047  $ & $    0.799^{+0.030}_{-0.027}$ & $    0.772^{+0.037}_{-0.029}$\\

\hline\hline                                                  
\end{tabular} 
}
\caption{Constraints at 68$\%$~CL errors on the cosmological parameters in case of the $12$ parameters model using different combinations of the datasets. The quoted upper/lower limits are at 95$\%$~CL.}
\label{tab12p}                                              
\end{table*}                                                    
\end{center}

\begin{figure*}[!hbtp]
\begin{center}
\includegraphics[width=.33\textwidth,keepaspectratio]{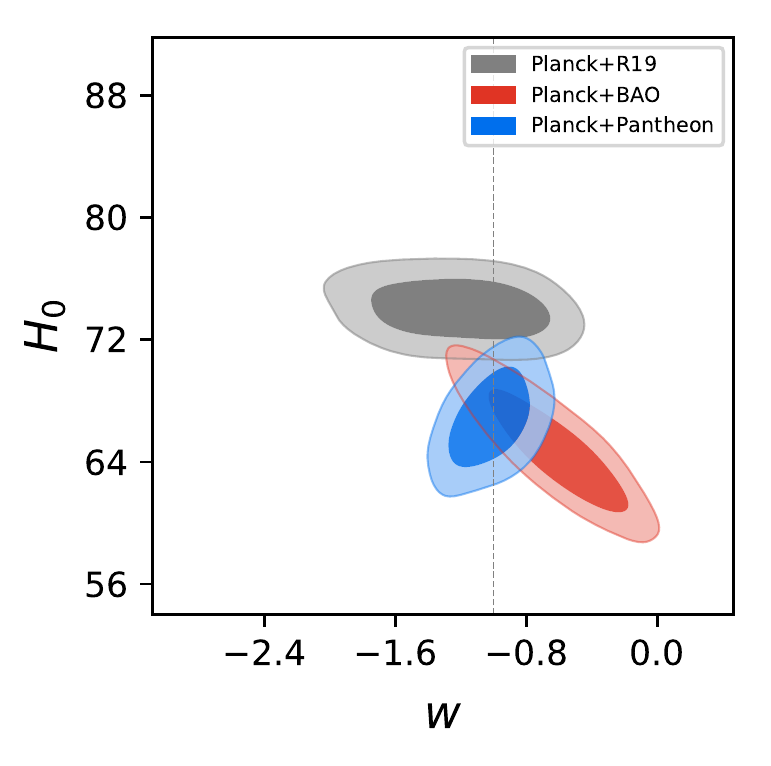}
\includegraphics[width=.33\textwidth,keepaspectratio]{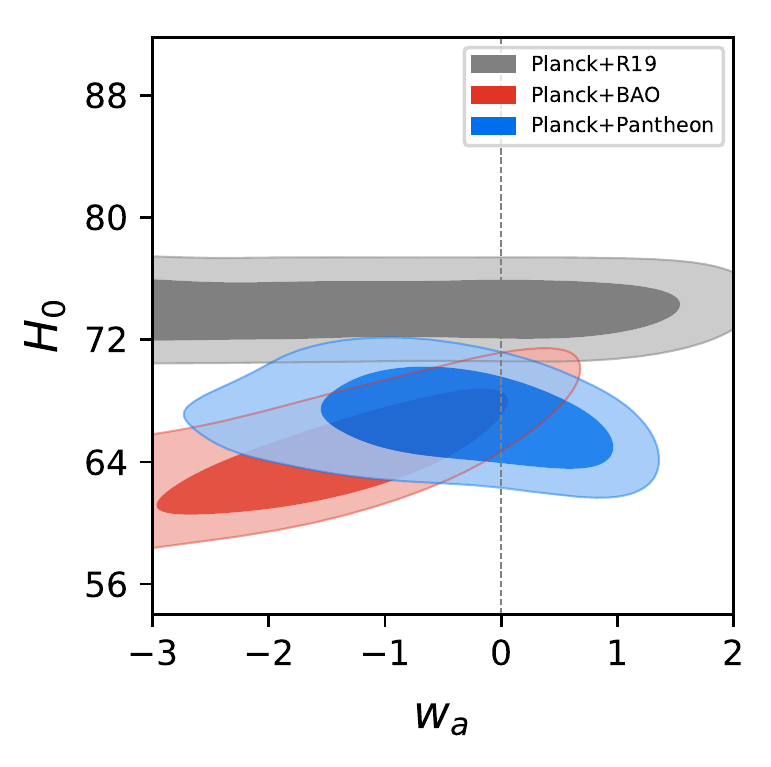}
\includegraphics[width=.33\textwidth,keepaspectratio]{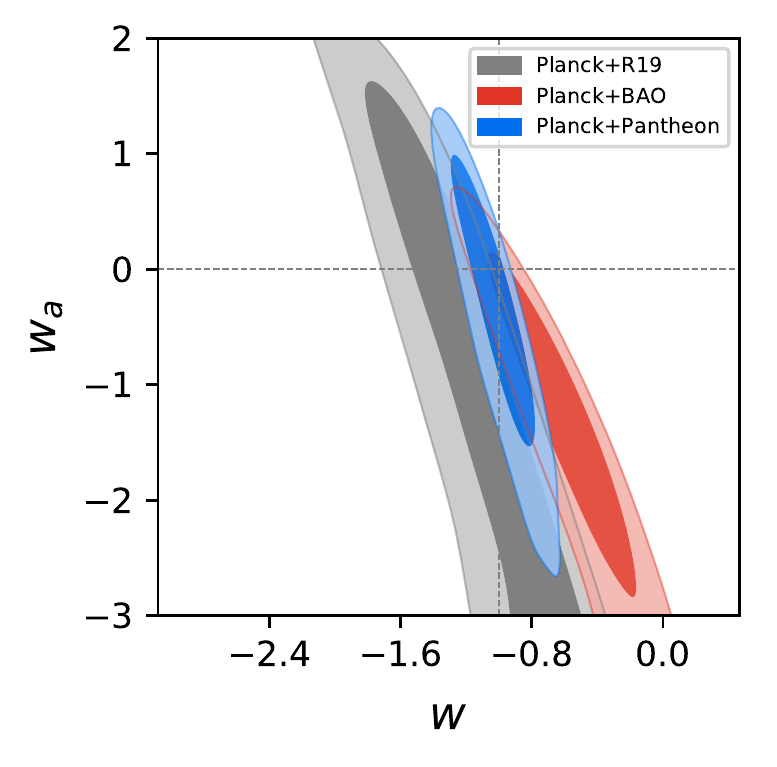}
\includegraphics[width=.33\textwidth,keepaspectratio]{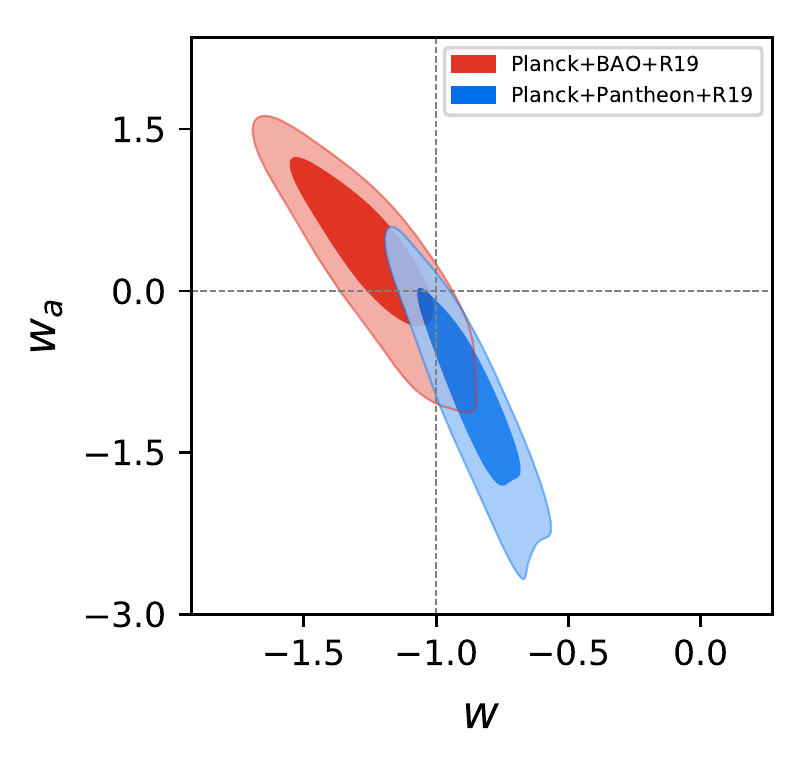}

\end{center}
\caption{Constraints at the $68\%$ and $95\%$ C.L. on the $w$ vs $H_0$,$w_a$ vs $H_0$ ,and $w$ vs $w_a$ planes for the Planck+Lensing, Planck+R19, Planck+BAO, and Planck+Pantheon datasets. In the bottom right corner we also show the 2D constraints
on the $w$ vs $w_a$ plane for the Planck+BAO+R19 and Planck+Pantheon+R19 datasets. A $12$ parameters model, $\Lambda$CDM+$w$+$w_a$+$\alpha_S$+$N_{eff}$+$\Sigma m_{\nu}$+$A_L$, is assumed in the analysis.}
\label{fig12p}
\end{figure*}

Let's now move to the $12$ parameter model. The constraints on the parameters from the different datasets are reported in Table~\ref{tab12p}. As one can see, the constraints on the density parameters $\Omega_bh^2$ and $\Omega_ch^2$, on the neutrino effective number $N_{eff}$, on $S_8$, and on inflationary parameters $n_s$ and $\alpha_s$, are almost identical to those obtained in the previous $11$ parameter case. These parameters are therefore only weakly affected by the introduction of a 
redshift-dependent dark energy equation of state. The introduction of $w_a$, however, has a significant impact on the constraints on the neutrino mass. Considering, for example, the Planck+BAO dataset, the bound on $\Sigma m_{\nu}$ is now practically a factor $\sim 1.46$ weaker with respect to the $11$ parameters case. A time-varying equation of state also increases the uncertainty on the Hubble constant from Planck+BAO. However,
$H_0$ from Planck+BAO  also shifts the Hubble constant towards lower values and a tension at the level of $3.2$ standard deviations is still present with the Planck+$R19$ dataset. The constraints on $H_0$ from Planck+Pantheon are less affected by the inclusion of $w_a$ and are almost identical to those obtained under the $11$ parameter scenario (with the exception of $w$ that strongly correlates with $w_a$).
Considering the $A_L$ parameter, we see that the evidence for $A_L>1$ is still present in the $12$ parameter case at between $2$ and $3$ standard deviations (with the exception of the Planck+lensing case where $A_L$ is consistent with the standard value to within two standard deviations).

It is interesting to consider the constraints on $w$ and $w_a$ and the correlation between these parameters. We do this in Figure~\ref{fig12p}, where we plot the 2-D constraints at the $68\%$ and $95\%$ C.L. on the $w$ vs $H_0$, $w_a$ vs $H_0$, and $w$ vs $w_a$ planes for the  Planck+R19, Planck+BAO, and Planck+Pantheon datasets. As already noted in \cite{DiValentino:2017zyq} in the case of the previous Planck 2015 release, when considering the constraints from Planck+R19 on the $w$ vs $w_a$ plane in Figure~\ref{fig12p} standard quintessence ($w>-1$ and $w_a>0$) and half of the "downward going" dark energy model space (characterized by an equation of state that decreases with time) are excluded at about $95 \%$ C.L.. The best-fit model for Planck+R19 has $w=-1.402$ and $w_a=-0.027$.
At the same time, Planck+BAO and Planck+Pantheon are consistent with a cosmological constant. While evolving dark energy is compatible with all of  the datasets, here is no evidence for $w_a \neq 0$ from any of them.

As we already discussed, in the $12$ parameter scenario, the tension between Planck+R19 and Planck+BAO and Planck+Pantheon is still present but reduced to the level of $\sim 3$ standard deviations. It is therefore interesting to investigate the constraints for the Planck+BAO+R19 and Planck+Pantheon+R19  datasets as we do in the bottom right panel of Figure~\ref{fig12p}. As one can see, the two combined datasets are also in tension: the Planck+BAO+R19  data prefers a solution with $w<-1$ and $w_a>0$ while the Planck+Pantheon+R19 points towards $w>-1$ and $w_a<0$. We can therefore argue that the CPL parametrization, while solving the Hubble tension between Planck and R19, still does not solve further tensions with Planck+BAO and Planck+Pantheon, and that these datasets itself are discordant when combined with R19.

\section{Conclusions}

In this paper, we have presented an updated analysis of the Planck 2018 Legacy data in an extended parameter space. First of all, we have found that the $A_L$ anomaly is still significantly present despite the increase in the number of parameters considered. This means that none of the extra parameters we included are able to fully describe this anomaly. The $A_L$ anomaly also significantly affects current bounds on the neutrino mass. Until the physical nature of this anomaly is clarified, current constraints on neutrino masses obtained under LCDM must be taken with a certain degree of scepticism. Secondly, we have shown that the current tension in the value of the Hubble constant between Planck 2018 and R19 can be solved by introducing a dark energy equation of state with $w<-1$ or such that $w_a<-6.66(1+w)$ when using the CPL parametrization for evolving dark energy. Quintessence model cannot therefore provide a solution to the Hubble tension. Moreover, a dark energy solution is preferred by the data relative to an increase in the neutrino effective number $N_{eff}$  that is, in our analysis, always consistent with the standard value $N_{eff}=3.046$.

\acknowledgments

EDV is supported from the European Research Council in the form of a Consolidator Grant 
with number 681431. AM thanks the University of Manchester and the Jodrell Bank Center for Astrophysics for hospitality.  AM is supported by TASP, iniziativa specifica INFN. 


\end{document}